\documentclass[a4paper]{article}

\usepackage{INTERSPEECH2021}

\usepackage{bm,amsmath,amssymb,mathrsfs,amsfonts}
\usepackage{caption}
\usepackage{scalefnt} %
\usepackage{multirow} %
\usepackage{here}
\usepackage{booktabs}
\usepackage{enumitem}
\usepackage{type1cm}
\usepackage{cite}
\usepackage{url}
\usepackage{algorithm}
\usepackage{algpseudocode}
\usepackage{arydshln}
\usepackage{color}

\newcommand{\deltadecot}{\delta}

\usepackage{pifont}%

\newcommand{\lambdactc}{\lambda_{\rm ctc}}
\newcommand{\lambdase}{\lambda_{\rm se}}
\newcommand{\lambdaqua}{\lambda_{\rm qua}}
\newcommand{\lambdalatency}{\lambda_{\rm latency}}

\newcommand{\lossctc}{\mathcal{L}_{\rm ctc}}
\newcommand{\lossqua}{\mathcal{L}_{\rm qua}}
\newcommand{\losslatency}{\mathcal{L}_{\rm latency}}
\newcommand{\lossrnnt}{\mathcal{L}_{\rm rnnt}}

\usepackage{etoolbox}

\newcommand{\algstrut}[1][\algruledefaultfactor]{\vrule width 0pt
depth .25\baselineskip height #1\baselineskip\relax}
\newcommand*{\algrule}[1][\algorithmicindent]{\hspace*{.5em}\vrule\algstrut
\hspace*{\dimexpr#1-.5em}}

\makeatletter
\newcount\ALG@printindent@tempcnta
\def\ALG@printindent{%
    \ifnum \theALG@nested>0%
    \ifx\ALG@text\ALG@x@notext%
    \else
    \unskip
    \ALG@printindent@tempcnta=1
    \loop
    \algrule[\csname ALG@ind@\the\ALG@printindent@tempcnta\endcsname]%
    \advance \ALG@printindent@tempcnta 1
    \ifnum \ALG@printindent@tempcnta<\numexpr\theALG@nested+1\relax%
    \repeat
    \fi
    \fi
}%

\patchcmd{\ALG@doentity}{\noindent\hskip\ALG@tlm}{\ALG@printindent}{}{\errmessage{failed to patch}}

\AtBeginEnvironment{algorithmic}{\lineskip0pt}

\title{StableEmit: Selection Probability Discount\\ for Reducing Emission Latency of Streaming Monotonic Attention ASR}

\name{Hirofumi Inaguma, Tatsuya Kawahara}
\address{
  Graduate School of Informatics, Kyoto University, Kyoto, Japan}
\email{\{inaguma,kawahara\}@sap.ist.i.kyoto-u.ac.jp}

\begin{document}

\maketitle
\begin{abstract}
While attention-based encoder-decoder (AED) models have been successfully extended to the online variants for streaming automatic speech recognition (ASR), such as monotonic chunkwise attention (MoChA), the models still have a large label emission latency because of the unconstrained end-to-end training objective. 
Previous works tackled this problem by leveraging alignment information to control the timing to emit tokens during training.
In this work, we propose a simple \textit{alignment-free} regularization method, \textit{StableEmit}, to encourage MoChA to emit tokens earlier.
StableEmit discounts the selection probabilities in hard monotonic attention for token boundary detection by a constant factor and regularizes them to recover the total attention mass during training.
As a result, the scale of the selection probabilities is increased, and the values can reach a threshold for token emission earlier, leading to a reduction of emission latency and deletion errors.
Moreover, StableEmit can be combined with methods that constraint alignments to further improve the accuracy and latency.
Experimental evaluations with LSTM and Conformer encoders demonstrate that StableEmit significantly reduces the recognition errors and the emission latency simultaneously.
We also show that the use of alignment information is complementary in both metrics.
\end{abstract}

\section{Introduction}
End-to-end (E2E) automatic speech recognition (ASR) models have been recently studied to streamline the complicated pipelines in conventional hybrid systems.
Online streaming is one of the most important tasks for real-world applications.
While frame-synchronous models such as connectionist temporal classification (CTC)~\cite{ctc_graves} and RNN transducer (RNN-T)~\cite{rnn_transducer} are natural choices for the streaming purpose, emerging progress of streaming attention-based encoder-decoder (AED) models, such as monotonic chunkwise attention (MoChA)~\cite{mocha}, make them good alternatives as well~\cite{inaguma2021alignment,kim2020review}.

Apart from the decoder topology, all streaming E2E models have a large emission latency because the alignment-unconstrained E2E training objective encourages the model to see as many future observations as possible.
To tackle this problem, various methods have been proposed; alignment path restriction~\cite{sak2015acoustic_asru,inaguma2020streaming,mahadeokar2020alignment}, encoder pre-training with frame-level supervision~\cite{hu2020exploring,inaguma2020streaming,mahadeokar2020alignment}, and a new training objective for direct latency minimization~\cite{inaguma2020streaming,inaguma2020_ctcsync,yu2021fastemit,yu2021dualmode,kim2021reducing}.
Among them, FastEmit~\cite{yu2021fastemit} successfully reduced the emission latency of RNN-T by modifying the loss to prioritize the generation of non-blank labels over blank labels \textit{without external alignment}.
However, the method is specific to RNN-T and cannot be applied to MoChA directly.
This is because MoChA detects token boundaries with a Bernoulli random variable in the internal attention module while RNN-T treats all labels, including a blank label, equally in the output softmax layer.

In this work, we propose a straightforward alignment-free regularization method, \textit{StableEmit}, to reduce the emission latency of MoChA.
StableEmit discounts selection probabilities for token boundary detection in MoChA with a constant factor during training.
Simultaneously, we encourage the expected attention weights calculated from the modified selection probabilities, to sum up to one to recover the attention mass (\textit{quantity regularization}~\cite{inaguma2020_ctcsync}).
Therefore, the confidence score for token emission is enhanced and reaches a threshold to emit a token earlier at test time.
This also reduces deletion errors, which stabilizes the search.

Moreover, it is possible to combine StableEmit with alignment regularization methods such as delay constrained training (DeCoT)~\cite{inaguma2020streaming} and CTC-synchronous training (CTC-ST)~\cite{inaguma2020_ctcsync}.
We further propose DeCoT-CTC, a new alignment path restriction method combining these ideas.

Experimental evaluations with Long Short-Term Memory (LSTM) and causal Conformer~\cite{gulati2020} encoders show that the proposed method significantly reduces the deletion errors and the emission latency of MoChA.
We also demonstrate that the combination with the alignment regularization methods has a complementary effect, leading to further improvements in the accuracy and latency.
The best MoChA model shows comparable performance to RNN-T on long-form speech utterances, which has been difficult to achieve so far.

\section{Background}\label{sec:background}

\subsection{Monotonic chunkwise attention (MoChA)}\label{ssec:mocha}
MoChA is a linear-time attention model on the basis of hard monotonic attention (HMA)~\cite{hard_monotonic_attention}.
To compensate for the performance degradation due to the hard attention to a single encoder output, MoChA introduces an additional soft attention  module restricted to a local chunk of $w$ frames from the boundary detected by HMA.
The test-time decoding complexity is linear, in which the discrete decision $z_{i,j}\in\{0,1\}$ to generate the $i$-th token on the $j$-th encoder output is sampled from a Bernoulli random variable parameterized by a selection probability $p_{i,j}$.

Because this procedure is not differentiable, the expected alignment probabilities $\alpha_{i,j}$ is calculated by summing all possible alignment probabilities during training as
\begin{eqnarray}
\alpha_{i,j} &=& p_{i,j}\sum_{k=1}^{j}\bigg(\alpha_{i-1,k}\prod_{l=k}^{j-1}(1-p_{i,l})\bigg) \nonumber \\
&=& p_{i,j}\bigg((1-p_{i,j-1})\frac{\alpha_{i,j-1}}{p_{i,j-1}}+\alpha_{i-1,j}\bigg). \label{eq:monotonic_attention_alpha}
\end{eqnarray}
While Eq.~\eqref{eq:monotonic_attention_alpha} requires the iterative update over $j$, there is an effective implementation which parallelizes the calculation by using the cumulative sum and product operations, denoted respectively as {\tt cumsum} and {\tt cumprod}, as
\scriptsize
\begin{eqnarray}
{\bm \alpha}_{i,:}={\bm p}_{i} \cdot \mbox{{\tt cumprod}}(1-{\bm p}_{i,:}) \cdot \mbox{{\tt cumsum}}\bigg(\frac{{\bm \alpha}_{i-1,:}}{\mbox{{\tt cumprod}}(1-{\bm p}_{i,:})}\bigg). \label{eq:monotonic_attention_alpha_parallel}
\end{eqnarray}
\normalsize
The chunkwise attention is estimated with $\alpha_{i,j}$ at training time and is calculated from the hard boundary at test time.
The interested readers are referred to~\cite{hard_monotonic_attention,mocha} for more details.

\subsection{Quantity regularization}\label{ssec:quantity_regularization}
An exponential decay of $\alpha_{i,j}$ due to the recurrence in Eq.~\eqref{eq:monotonic_attention_alpha} enlarges the mismatch between training and testing behaviors, which is the main cause of the poor performance of the original MoChA~\cite{online_hybrid_ctc_attention,inaguma2021alignment,online_hybrid_ctc_attention_taslp2020}.
To mitigate this problem, quantity regularization~\cite{inaguma2020streaming} was proposed by introducing a quantity loss $\lossqua$, which makes the expectation of the total number of boundaries closer to the output sequence length $U$ as $\lossqua = |U - \sum_{i,j}{\alpha_{i,j}}|.$

\subsection{Alignment path restriction}\label{ssec:alignment_path_restriction}
One of the effective methods to reduce the emission latency of MoChA is to ignore paths in Eq.~\eqref{eq:monotonic_attention_alpha} by referring to external alignment information extracted from the hybrid system~\cite{inaguma2020streaming}.
Given reference boundaries ${\bf b}^{\rm ref}=({\rm b}^{\rm ref}_{1}, \cdots, {\rm b}^{\rm ref}_{U})$ estimated from the word alignment, delay constrained training (DoCoT)~\cite{inaguma2020streaming} masks out inappropriate paths based on
\begin{eqnarray*}
\alpha_{i,j} = \begin{cases}
     p_{i,j}\bigg((1-p_{i,j-1})\frac{\alpha_{i,j-1}}{p_{i,j-1}}+\alpha_{i-1,j}\bigg) & (j \le {\rm b}^{\rm ref}_{i} + \delta) \\
     0 & ({\rm otherwise}),
\end{cases}
\end{eqnarray*}
where $\delta$ is a hyperparameter to control the delay.
To recover the reduced scale of $\alpha_{i,j}$, it is essential to use quantity regularization~\cite{inaguma2020streaming}.

Moreover, it is also possible to use CTC alignment to estimate $\bf{b}^{\rm ref}$.
We refer to this method as \textit{DeCoT-CTC} and note that it is newly investigated in this work.
Although the CTC alignment is not guaranteed to be accurate, it will be effective because the model still has some freedom to learn the optimal timing to emit tokens, unlike point-wise boundary restriction in the expected latency loss described in Section~\ref{ssec:expected_latency_loss}.
Moreover, the CTC alignment can be obtained without frame-level alignment supervision.
During training, we use the most probable CTC alignment as in~\cite{inaguma2020_ctcsync}.

\subsection{Expected latency loss}\label{ssec:expected_latency_loss}
Another strategy to reduce the latency is to use boundary supervision by minimizing the expected latency loss~\cite{inaguma2020streaming,inaguma2020_ctcsync} as
\begin{eqnarray}
{\rm b}^{\rm mocha}_{i} &=& \sum_{j=1}^{T'}j \cdot \alpha_{i,j}, \nonumber \\
\losslatency &=& \frac{1}{U}\sum_{i=1}^{U}{|{\rm b}^{\rm ref}_{i} - {\rm b}_{i}^{\rm mocha}|}, \nonumber
\end{eqnarray}
where ${\rm b}^{\rm mocha}_{i}$ is the expected boundary index for the $i$-th token, and $T'$ is the time length of encoder outputs.
When using the alignment from a hybrid system, training becomes minimum latency training (MinLT)~\cite{inaguma2020streaming}.
When using the CTC alignment, it becomes CTC-synchronous training (CTC-ST)~\cite{inaguma2020_ctcsync}.
We have shown that CTC-ST outperforms MinLT in the accuracy and latency~\cite{inaguma2021alignment} and therefore focus on the former in this work.

\subsection{Total training objective}
To summarize the above methods, the generalized training objective of MoChA is formulated as follows:
\begin{multline}
\mathcal{L}^{\rm total}_{\rm mocha} = (1- \lambdactc) \mathcal{L}_{\rm mocha} + \lambdactc \lossctc \\ 
+ \lambdalatency \losslatency + \lambdaqua \lossqua, \label{eq:total_loss}
\end{multline}
where $\lambda_{*}$ is a corresponding task weight.
When $\lambdalatency > 0$, $\lambdaqua$ is set to 0.0 in this work~\cite{inaguma2020_ctcsync}.
We summarize the above alignment regularization methods in Table~\ref{tab:mocha_alignment_summary}.

\begin{table}[t]
    \centering
    \begingroup
    \caption{Summary of alignment regularization for MoChA}\label{tab:mocha_alignment_summary}
    \vspace{-2mm}
    \scalebox{1.0}{
    \begin{tabular}{c|l|c|c} \toprule
        \multicolumn{2}{c}{} & \multicolumn{2}{c}{Source of alignment} \\ \cmidrule(lr){3-4}
        &  & CTC & Hybrid ASR \\
        \midrule
        \multirow{4}{*}{\rotatebox{90}{Training}} & \multirow{2}{*}{\shortstack{Expected latency\\loss}} & \multirow{2}{*}{CTC-ST~\cite{inaguma2020_ctcsync}} & \multirow{2}{*}{MinLT~\cite{inaguma2020streaming}} \\
        &  &  &  \\ \cmidrule(lr){2-4}
        
       & \multirow{2}{*}{\shortstack{Alignment path\\restriction}} & \multirow{2}{*}{\shortstack{DeCoT-CTC\\(this work)}} & \multirow{2}{*}{DeCoT~\cite{inaguma2020streaming}} \\
       &  &  &  \\
      \bottomrule
    \end{tabular}
    }
    \endgroup
    \vspace{-2mm}
\end{table}

\section{StableEmit}\label{sec:proposed_method}
In this section, we propose \textit{StableEmit}, a simple but effective emission latency regularization method for MoChA without external alignment information.

\subsection{Motivation}
While the methods in Section~\ref{ssec:alignment_path_restriction} and Section~\ref{ssec:expected_latency_loss} can reduce the emission latency of MoChA, they require alignment information, although the CTC alignment can be obtained easily.
Moreover, at test time, token boundaries are detected when and only when $p_{i,j}$ surpasses a threshold $\tau=0.5$~\cite{hard_monotonic_attention}.\footnote{Setting $\tau=0.5$ is reasonable when assuming $p_{i,j}$ is well discretized after the training.}
Therefore, if the confidence of boundary prediction in some speech frames is below $\tau$, the corresponding token cannot be generated because it does not appear in the subsequent search process.
As a result, this leads to increasing deletion errors, especially for long-form speech data.

\subsection{Formulation}
To solve this problem, we propose \textit{StableEmit}, which discounts $p_{i,j}$ \textit{during training only} to detect token boundaries earlier with high confidence at test time.
Specifically, we introduce a constant discount factor $\lambdase$ $(0 \leq \lambdase < 1)$ and multiply $p_{i,j}$ by $1 - \lambdase$.
Therefore, Eq.~\eqref{eq:monotonic_attention_alpha} is reformulated as
\begin{eqnarray}
p'_{i,j} &=& (1 - \lambdase) \cdot p_{i,j}, \\ \label{eq:stable_emit}
\alpha_{i,j} &=& p'_{i,j}\bigg((1-p'_{i,j-1})\frac{\alpha_{i,j-1}}{p'_{i,j-1}}+\alpha_{i-1,j}\bigg). \nonumber
\end{eqnarray}
Accordingly, Eq.~\eqref{eq:monotonic_attention_alpha_parallel} is rewritten as
\footnotesize
\begin{eqnarray*}
{\bm \alpha}_{i,:}={\bm p}'_{i,:} \cdot \mbox{{\tt cumprod}}(1-{\bm p}'_{i,:}) \cdot \mbox{{\tt cumsum}}\bigg(\frac{{\bm \alpha}_{i-1,:}}{\mbox{{\tt cumprod}}(1-{\bm p}'_{i,:})}\bigg).
\end{eqnarray*}
\normalsize
Similarly to FastEmit, StableEmit is also a \textit{sequence-level} regularization because the discount of $p_{i,j}$ at every grid $(i,j)$ influences $\alpha_{i,j}$ globally.
As the discount decreases $\sum_{j} \alpha_{i,j}$, we further need to use the quantity regularization to recover it.
With this technique, $p_{i,j}$ is optimized to detect token boundaries with high confidence because it is encouraged to compensate the discount so that $\sum_{j} \alpha_{i,j}$ gets closer to 1.
In other words, the reasonable value of $\tau$ increases from 0.5 to $0.5 / (1 - \lambdase)$ through the training.
However, at test time, we keep $\tau = 0.5$ as in the original formulation, unless otherwise noted.
This is effective for reducing the emission latency because $p_{i,j}$ can reach $\tau$ earlier in terms of $j$ if it is trained to recover the discount properly.

Since StableEmit just modifies $p_{i,j}$ during training, it can be combined with alignment regularization methods that manipulate $\alpha_{i,j}$ such as DeCoT and CTC-ST.
The effect of combination will be described in Section~\ref{sec:experiment}.

\section{Experimental evaluations}\label{sec:experiment}
\subsection{Experimental setup}\label{ssec:setup}
We used the TEDLIUM release v2 (TEDLIUM2)~\cite{tedlium}, a 210-hour English lecture speech corpus, and the evaluation sets include long-form utterances (40/30 seconds in the \textit{dev}/\textit{test} sets, respectively).
For the GPU memory capacity, however, we removed utterances longer than 16 and 20 seconds in the training data for LSTM and Conformer models, respectively.
We extracted 80-channel log-mel filterbank coefficients computed with a 25-ms window that was shifted every 10ms with Kaldi~\cite{kaldi}.
We applied three-fold speed perturbation~\cite{speed_perturbation} and SpecAugment~\cite{specaugment}.
The vocabularies were constructed by the byte pair encoding (BPE) algorithm ~\cite{sennrich2015neural} with 10k and 1k units for AED and RNN-T models, respectively.

For streaming encoders, we used the unidirectional LSTM (UniLSTM) and causal Conformer (UniConformer)~\cite{gulati2020} architectures, both of which follow four CNN layers with a $3 \times 3$ filter.
A max-pooling layer with a stride of $2 \times 2$ was inserted at the 2nd and 4th CNN layer for LSTM and the 4th CNN layer for UniConformer.
We used five layers of UniLSTM encoder with 1024 units~~\cite{inaguma2021alignment}.
We adopted Conformer (M) configuration in~\cite{gulati2020} while reducing the number of blocks from 16 to 12.
Moreover, we replaced batch normalization~\cite{ioffe2015batch} in each convolution module with layer normalization~\cite{ba2016layer}.\footnote{We also tried group normalization~\cite{wu2018group}, but layer normalization was the best in our implementations.}
We used Shaw's style relative positional encoding (RPE)~\cite{shaw-etal-2018-self} and clipped the relative distance to 10 in each block.
For streaming, we adopted causal depthwise separable convolution with a kernel size of 7 and a causal self-attention mask.
Therefore, the lookahead latency was introduced in the frontend CNN layers only.
Furthermore, we used hierarchical downsampling~\cite{dong2019self}, where the max-pooling with a stride of 2 was performed in the last frontend CNN layer, 4th, and 8th Conformer blocks, respectively.
This resulted in the downsampling factor of 8 for the UniConformer in total while 4 for the LSTM.
The total lookahead latency in the UniLSTM and UniConformer encoders were 60ms and 40ms, respectively.
The decoder consisted of a single layer of LSTM with 1024 units for MoChA, with a chunk size $w = 4$.
For RNN-T, we used a two-layer LSTM prediction network with 1024 units and a joint network with 512 units.

The Adam algorithm~\cite{adam} was used for optimization.
We set $\lambdactc$ to 0.3 in all models, including RNN-T.\footnote{RNN-T is optimized via $\mathcal{L}^{\rm total}_{\rm rnnt} =(1- \lambdactc) \lossrnnt + \lambdactc \lossctc$.}
We set ($\lambdalatency$, $\lambdaqua$) to (1.0, 0.0) for CTC-ST, otherwise (0.0, 2.0).
During inference, we use the beam width $b$ of 10 with a four-layer LSTM language model having 1024 units.
We used length normalization~\cite{murray-chiang-2018-correcting} and softmax smoothing with a temperature of 0.7~\cite{chorowski2017towards}.
Our implementation is publicly available.\footnote{\url{https://github.com/hirofumi0810/neural_sp}.}

\begin{table}[t]
    \centering
    \begingroup
    \caption{Results on TEDLIUM2. Token emission latency (TEL) is calculated on the test set. ${}^{\spadesuit}$SpecAugment was NOT used.
    }\label{tab:result_tedlium2}
    \vspace{-3mm}
    \scalebox{0.85}{
    \begin{tabular}{lccccc} \toprule
     \multirow{2}{*}{Model} & \multicolumn{2}{c}{WER [$\%$] ($\downarrow$)} & \multicolumn{3}{c}{TEL [ms] ($\downarrow$)} \\ \cmidrule(lr){2-3} \cmidrule(lr){4-6}
     \multicolumn{1}{c}{} & dev & test & 50th & 90th & 95th \\ \midrule
     BiLSTM - Global AED & \phantom{0}8.1 & \phantom{0}7.5 & -- & -- & -- \\
     
     UniLSTM - RNN-T ($b=5$) & 11.0 & 11.0 & -- & -- & -- \\ %
     UniLSTM - RNN-T & 10.7 & 10.7 & -- & -- & -- \\ %
     \hline
     UniLSTM - MoChA${}^{\spadesuit}$ & 16.3 & 14.9 & 280 & 680 & 1120 \\ 
     \ + StableEmit${}^{\spadesuit}$ & \bf{13.9} & \bf{12.7} & \bf{240} & \bf{440} & \phantom{0}\bf{640} \\

     \ \ \ + DeCoT-CTC ($\deltadecot=12$) & \bf{12.4} & \bf{10.7} & \bf{240} & \bf{400} & \phantom{0}\bf{480} \\
     \ \ \ + DeCoT-CTC ($\deltadecot=16$) & 12.5 & 11.1 & 280 & 440 & \phantom{0}520 \\
     \hline
     \ + DeCoT ($\deltadecot=16$) & 12.4 & \bf{10.7} & 280 & 440 & \phantom{0}560 \\
     \ \ \ + StableEmit & \bf{12.0} & \bf{10.7} & \bf{240} & \bf{360} & \phantom{0}\bf{440} \\ \hline
     \ + CTC-ST & 13.5 & 11.3 & 200 & 360 & \phantom{0}480 \\
     \ \ \ + StableEmit & \bf{11.7} & \bf{10.9} & 200 & \bf{320} & \phantom{0}\bf{440} \\
     \midrule \midrule
    
     Conformer - Global AED & \phantom{0}7.3 & \phantom{0}7.0 & -- & -- & -- \\
    
     UniConformer - CTC & 12.1 & 11.2 & 240 & 320 & \phantom{0}400 \\
     UniConformer - RNN-T & \phantom{0}8.5 & \phantom{0}8.2 & -- & -- & -- \\ \hline %
     
     UniConformer - MoChA & \phantom{0}9.6 & \phantom{0}8.8 & 320 & 480 & \phantom{0}640 \\
     \ + StableEmit & \phantom{0}\bf{9.1} & \phantom{0}\bf{8.4} & \bf{240} & 480 & \phantom{0}\bf{480} \\
     \ \ \ + DeCoT-CTC ($\deltadecot=8$) & \phantom{0}\bf{8.8} & \phantom{0}\bf{8.4} & \bf{240} & \bf{400} & \phantom{0}\bf{480} \\
     \ \ \ + DeCoT-CTC ($\deltadecot=12$) & \phantom{0}\bf{8.5} & \phantom{0}\bf{8.4} & 320 & 480 & \phantom{0}\bf{480} \\
     \hline
     
     \ + DeCoT ($\deltadecot=8$) & \phantom{0}8.5 & \phantom{0}8.5 & \bf{240} & 400 & \phantom{0}480 \\
     \ \ \ + StableEmit & \phantom{0}8.7 & \phantom{0}8.7 & \bf{240} & \bf{320} & \phantom{0}\bf{400} \\
     \ + DeCoT ($\deltadecot=12$) & \phantom{0}8.7 & \phantom{0}8.5 & 320 & 480 & \phantom{0}560 \\
     \ \ \ + StableEmit & \phantom{0}\bf{8.6} & \phantom{0}\bf{8.2} & \bf{240} & \bf{400} & \phantom{0}\bf{480} \\ \hline
     \ + CTC-ST & \phantom{0}\bf{8.9} & \phantom{0}8.8 & \bf{240} & \bf{320} & \phantom{0}\bf{400} \\
     \ \ \ + StableEmit & \phantom{0}9.3 & \phantom{0}\bf{8.5} & \bf{240} & \bf{320} & \phantom{0}\bf{400} \\
     \bottomrule
    \end{tabular}
    }
    \endgroup
    \vspace{-4mm}
\end{table}

\subsection{Main results}\label{ssec:main_result}
\vspace{-1mm}
We summarized the results in Table~\ref{tab:result_tedlium2}.
We used forced alignment results to evaluate the token emission latency (TEL)~\cite{inaguma2021alignment} from the ground-truth acoustic boundary.
We report median, 90th, and 95th percentile latency calculated on the \textit{test} set.

\vspace{-1mm}
\subsubsection{UniLSTM encoder models}
\vspace{-1mm}
We first confirmed significant improvement by StableEmit over the baseline UniLSTM MoChA when SpecAugment was not used.
We activated SpecAugment with alignment regularization methods after pre-training the baseline model to stabilize the training~\cite{inaguma2020_ctcsync,inaguma2020streaming}.
While both CTC-ST and DeCoT were effective, the combination with StableEmit brought further improvements with an additional TEL reduction.
We confirmed that DeCoT-CTC was also effective with a smaller $\deltadecot$ than that in DeCoT because CTC also had a delay.
Specifically, applying StableEmit to CTC-ST showed relative WER improvements of 13.3\% (\textit{dev}) and 3.5\% (\textit{test}), and absolute TEL reductions of 40ms and 80ms in the 90th and 95th percentile, respectively.
Compared to the baseline model, the TEL was reduced by 80ms, 160ms, and 240ms in the 50th, 90th, and 95th percentile, respectively.
We also observed that StableEmit was effective in reducing the tail of TEL distributions for DeCoT.

\begin{table}[t]
    \centering
    \begingroup
    \caption{Effects of boundary threshold and loss weight on TEDLIUM2. The encoder is UniConformer.}\label{tab:threshold_stableemit}
    \vspace{-3mm}
    \scalebox{0.82}{
    \begin{tabular}{ccccccc} \toprule
      \multirow{2}{*}{$\lambdase$} & \multirow{2}{*}{$\tau$} & \multicolumn{2}{c}{WER [$\%$] ($\downarrow$)} & sub / ins / del & \multicolumn{2}{c}{TEL [ms] ($\downarrow$)} \\ \cmidrule(lr){3-4} \cmidrule(lr){5-5} \cmidrule(lr){6-7}
      &  & dev & test & dev & 50th & 90th \\ 
      \midrule
      
      \multirow{5}{*}{0.0\phantom{0}} & 0.5 & \phantom{0}9.58 & 8.76 & 4.88 / 1.55 / 3.15 & 320 & 480 \\
      & 0.4 & \phantom{0}\bf{9.50} & 8.48 & 4.96 / 1.77 / 2.77 & 320 & 480 \\
      & 0.3 & \phantom{0}9.64 & 8.39 & 4.95 / 2.16 / 2.52 & 320 & 480 \\
      & 0.2 & 10.11 & \bf{8.32} & 5.00 / 2.79 / 2.31 & 320 & 480 \\
      & 0.1 & 11.70 & 8.70 & 5.00 / 4.78 / 1.91 & \bf{240} & \bf{400} \\
      \midrule
      
      0.0* & 0.5 & \phantom{0}9.80 & 9.00 & 4.92 / 1.45 / 3.43 & 320 & 560 \\
      \midrule \midrule
      
      \multirow{5}{*}{0.1\phantom{0}} & 0.5 & \phantom{0}9.13 & 8.45 & 4.94 / 1.48 / 2.71 & 240 & 480 \\
      & 0.4 & \phantom{0}9.10 & 8.47 & 4.93 / 1.57 / 2.60 & 240 & \bf{400} \\ 
      & 0.3 & \phantom{0}\bf{9.06} & 8.39 & 5.06 / 1.61 / 2.40 & 240 & \bf{400} \\
      & 0.2 & \phantom{0}9.12 & \bf{8.29} & 5.16 / 1.77 / 2.19 & 240 & \bf{400} \\
      & 0.1 & \phantom{0}9.63 & 8.45 & 5.26 / 2.28 / 2.15 & 240 & \bf{400} \\ \midrule
      
      0.1* & 0.5 & \phantom{0}9.13 & 8.55 & 4.97 / 1.42 / 2.74 & 240 & 480 \\
      \midrule \midrule
      
      0.05 & \multirow{3}{*}{0.5} & \phantom{0}9.15 & 8.56 & 4.92 / 1.51 / 2.73 & 320 & 480 \\
      0.15 &  & \phantom{0}9.38 & 8.45 & 4.90 / 1.73 / 2.76 & \bf{240} & \bf{400} \\
      0.2\phantom{0} &  & 10.01 & 8.99 & 4.86 / 2.47 / 2.67 & \bf{240} & \bf{400} \\
      \bottomrule
    \end{tabular}
    }
    \vspace{-2mm}
    \begin{flushright}
    {*$\lambdase$ was set to 0.1 at test time.}
    \end{flushright}
    \endgroup
    \vspace{-4mm}
\end{table}

\vspace{-1mm}
\subsubsection{UniConformer encoder models}
\vspace{-1mm}
Next, we applied StableEmit to the UniConformer models.
Unlike the UniLSTM models, we activated SpecAugment from scratch and activated $\losslatency$ and $\lossqua$ after running for 10 epochs, without pre-training.
Firstly, we observed a significant improvement from the UniLSTM models with the Conformer encoder.
StableEmit was also effective in this case, leading to relative WER improvements of 5.2\% (\textit{dev}) and 4.5\% (\textit{test}) and absolute reduction of the median TEL by 80ms.
Leveraging alignment information further reduced the TEL and WER in some cases.
In this regard, we observed that alignment path restriction (DeCoT and DeCoT-CTC) was more effective than CTC-ST in terms of WER while CTC-ST reached the same level of TEL as the pure CTC model without external alignment.
Although the gains became smaller than those of the UniLSTM models, we attributed it to the larger downsampling factor, which improved the baseline performance significantly, as shown in Section~\ref{ssec:ablation}.

\vspace{-1mm}
\subsection{Analysis}\label{ssec:analysis}
\vspace{-1mm}
Next, we study the effect of a discount factor $\lambdase$ (for training) and a threshold $\tau$ for selection probabilities (for testing).
We used the UniConformer encoder for this purpose.
The results in Table~\ref{tab:threshold_stableemit} showed that reducing $\tau$ from 0.5 slightly reduced WER regardless of the use of StableEmit.
However, the model trained with StableEmit consistently outperformed the baseline in both WER and TEL and was robust to a small $\tau$.
Furthermore, we also set $\lambdase$ to 0.1 during testing, but this did not improve WER nor TEL.
This verifies that the confidence of $p_{i,j}$ at test time is increased by StableEmit.
For simplicity, we keep $\tau = 0.5$ in all other experiments.

We also investigate the effect of different values of $\lambdase$.
Increasing $\lambdase$ reduced the 90th percentile TEL by 80ms while at the cost of WER.
Regarding error types, StableEmit successfully reduced the deletion errors while a strong constraint by a small $\tau$ or a large $\lambdase$ resulted in increasing the insertion errors.

\begin{table}[t]
    \centering
    \begingroup
    \caption{Ablation study on TEDLIUM2}\label{tab:result_ablation}
    \vspace{-3mm}
    \scalebox{0.82}{
    \begin{tabular}{lcc} \toprule
      \multirow{2}{*}{Model} & \multicolumn{2}{c}{WER [$\%$] ($\downarrow$)} \\ \cmidrule(lr){2-3}
       & dev & test  \\ \midrule
      UniConformer - MoChA + StableEmit & \phantom{0}\bf{9.1} & \phantom{0}\bf{8.4} \\
      \ \ w/o quantity reg. ($\lambdaqua=0$ in Eq.~\eqref{eq:total_loss}) & \phantom{0}9.7 & \phantom{0}9.9 \\
      \ \ w/o StableEmit ($\lambdase=0$ in Eq.~\eqref{eq:stable_emit}) & \phantom{0}9.6 & \phantom{0}8.8 \\
      \ \ w/o hierarchical downsampling & 10.3 & 10.0 \\
      \ \ \ \ w/o StableEmit & 11.1 & 10.4 \\
      \ \ w/o CTC loss ($\lambdactc=0$ in Eq.~\eqref{eq:total_loss}) & \phantom{0}9.2 & \phantom{0}8.8 \\
      \ \ LayerNorm $\to$ BatchNorm & 31.3 & 23.4 \\
      \midrule
      UniConformer - RNN-T & \phantom{0}\bf{8.5} & \phantom{0}\bf{8.2} \\
      \ \ w/o hierarchical downsampling & \phantom{0}8.7 & \phantom{0}8.4 \\
      \ \ w/o CTC loss ($\lambdactc=0$) & \phantom{0}8.7 & \phantom{0}8.6 \\
      \ \ LayerNorm $\to$ BatchNorm & \phantom{0}8.8 & \phantom{0}\bf{8.2} \\
      \bottomrule
    \end{tabular}
    }
    \endgroup
    \vspace{-2mm}
\end{table}

\vspace{-1mm}
\subsection{Ablation study}\label{ssec:ablation}
\vspace{-1mm}
We also conduct the ablation study in Table~\ref{tab:result_ablation}.
We observed that the quantity regularization was essential for StableEmit.
The hierarchical downsampling and layer normalization were also key ingredients when the UniConfomer encoder was used.
Moreover, we found that the hierarchical downsampling, the auxiliary CTC loss, and layer normalization were also beneficial for RNN-T.

\begin{figure}[t]
  \centering
  \begingroup
  \includegraphics[width=1.0\linewidth]{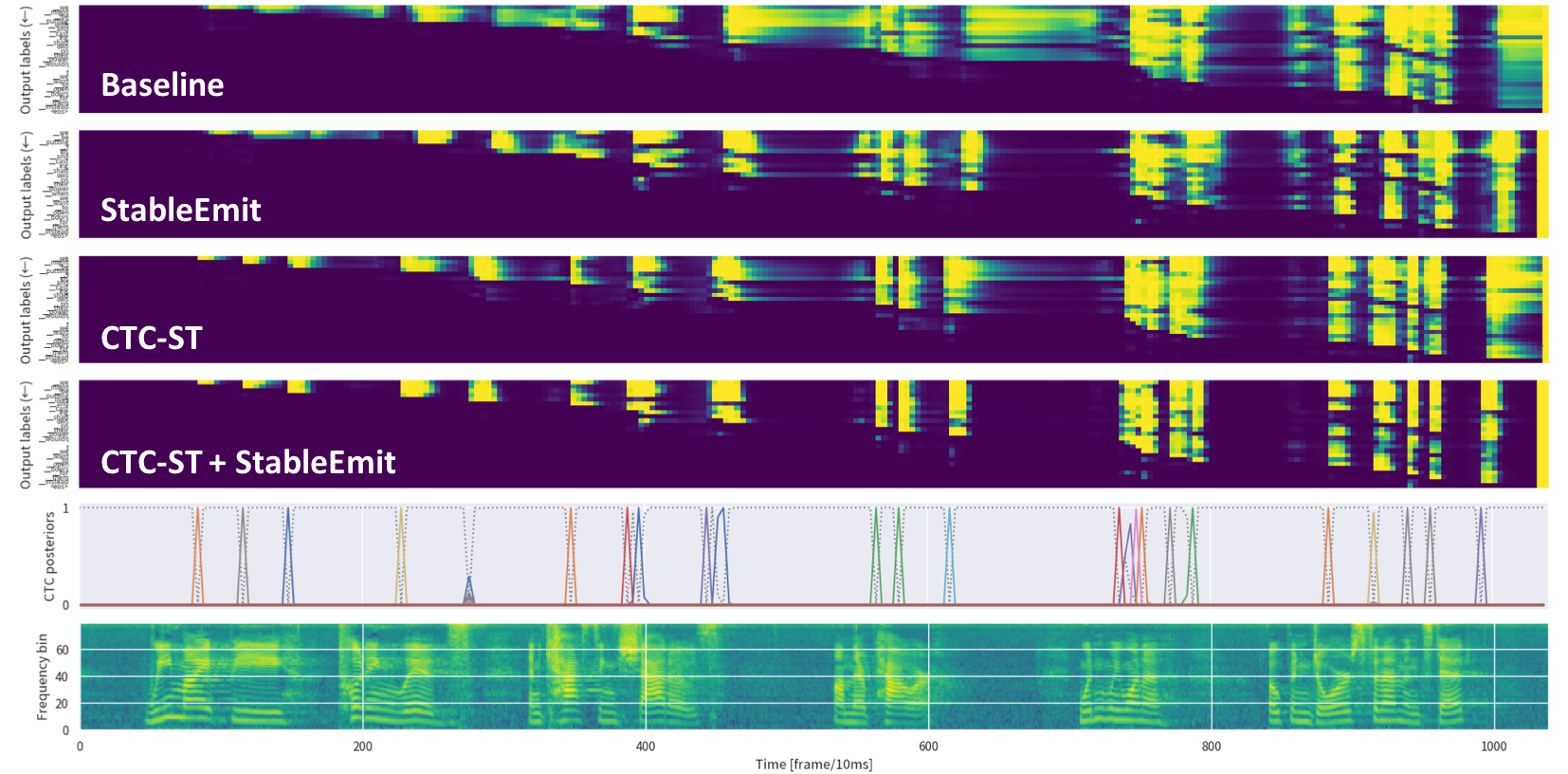}
  \vspace{-5mm}
  \caption{Visualization of selection probabilities and CTC outputs from UniLSTM MoChA and the input spectrogram. Yellow regions represent high probabilities.}
  \label{fig:visualization_p_choose}
  \endgroup
  \vspace{-1mm}
\end{figure}

\vspace{-1mm}
\subsection{Visualization of selection probabilities}\label{ssec:visualization_selection_prob}
\vspace{-1mm}
We visualize the selection probabilities $p_{i,j}$ in the UniLSTM MoChA models in Figure~\ref{fig:visualization_p_choose}.
The yellow regions represent high probabilities close to 1.
We used the whole encoder outputs for visualization.
We also plotted the CTC probabilities obtained from the model trained with CTC-ST and StableEmit.
We can see that the baseline MoChA had blurred probabilities, and StableEmit sharpened the distribution, indicating the confidence to emit a token was increased.
CTC-ST shifted some token boundaries to the left and sharpened the probability distribution~\cite{inaguma2021alignment}.
Moreover, its combination with StableEmit further made them clearer, showing a complementary effect.
The delay became almost the same as that of CTC.

\begin{table}[t]
    \centering
    \begingroup
    \caption{Results on Librispeech. TEL was averaged over the test-clean and test-other sets.}\label{tab:result_librispeech}
    \vspace{-3mm}
    \scalebox{0.78}{
    \begin{tabular}{lcccccc} \toprule
     \multirow{3}{*}{Model} & \multicolumn{4}{c}{WER [$\%$] ($\downarrow$)} & \multicolumn{2}{c}{TEL [ms] ($\downarrow$)} \\ \cmidrule(lr){2-5} \cmidrule(lr){6-7}
     \multicolumn{1}{c}{} & \multicolumn{2}{c}{dev} & \multicolumn{2}{c}{test} & \multirow{2}{*}{50th} & \multirow{2}{*}{90th} \\
     \multicolumn{1}{c}{} & clean & other & clean & other &  &   \\ \midrule
     UniConformer - RNN-T  & 2.8 & 8.2 & 3.1 & 8.5 & -- & -- \\
     UniConformer - MoChA & 2.6 & 7.6 & 2.8 & 7.9 & 400 & 560 \\
     \ \ + StableEmit & 2.9 & \bf{7.5} & 3.0 & 8.2 & \bf{320} & \bf{480} \\
     \bottomrule
    \end{tabular}
    }
    \endgroup
    \vspace{-4mm}
\end{table}

\vspace{-1mm}
\subsection{Results on Librispeech}
\vspace{-1mm}
Finally, we present results on Librispeech (960h)~\cite{librispeech} in Table~\ref{tab:result_librispeech}.
We used the same architecture as previous experiments.
We used $\lambdaqua=0.1$ for the baseline and $(\lambdaqua$, $\lambdase$) = (0.2, 0.2) for StableEmit.
We observed TEL reductions of 80ms in the 50th and 90th percentile at the cost of a slight WER degradation.
This is because MoChA did not suffer from the deletion errors on this corpus.
However, it outperformed a strong RNN-T model having the same encoder architecture.

\section{Conclusions}\label{sec:conclusion}
In this work, we have proposed StableEmit, a simple alignment-free solution to reduce the emission latency of MoChA for streaming ASR applications.
We encouraged the model to learn high selection probabilities in the token boundary detection by discounting them during training as a regularization.
We experimentally demonstrated that StableEmit reduced not only the emission latency but also the deletion errors.
Combining it with the alignment regularization further reduced the latency.

\bibliographystyle{IEEEtran}
\bibliography{reference}

\end{document}